\documentclass[11pt,a4paper]{article}
\usepackage[dvips,colorlinks,bookmarksopen,bookmarksnumbered,linkcolor=black,citecolor=black,urlcolor=black,breaklinks]{hyperref}
\usepackage{graphicx}
\usepackage{breakurl}

\begin{document}

\title{A confidence interval robust to publication bias for random-effects meta-analysis of few studies}

\author{M.~Henmi$^1$, S.~Hattori$^2$, T.~Friede$^3$\thanks{\textit{correspondence to}: Tim Friede, Department of Medical Statistics, University Medical Center G\"{o}ttingen, Germany; email: \texttt{tim.friede@med.uni-goettingen.de}} \\ \\
\footnotesize $^1$Institute of Statistical Mathematics, 10-3 Midori-cho, Tachikawa, Tokyo 190-8562, Japan\\
\footnotesize $^2$Department of Biomedical Statistics, Graduate School of Medicine, Osaka University, Osaka 565-0871, Japan\\
\footnotesize $^3$Department of Medical Statistics, University Medical Center G\"ottingen, G\"ottingen, Germany}

\date{}

\maketitle

\begin{abstract}
Systematic reviews aim to summarize all the available evidence relevant to a particular research question. If appropriate, the data from identified studies are quantitatively combined in a meta-analysis. Often only few studies regarding a particular research question exist. In these settings the estimation of the between-study heterogeneity is challenging. Furthermore, the assessment of publication bias is difficult as standard methods such as visual inspection or formal hypothesis tests in funnel plots do not provide adequate guidance. Previously, Henmi and Copas (Statistics in Medicine 2010, 29: 2969--2983) proposed a confidence interval for the overall effect in random-effects meta-analysis that is robust to publication bias to some extent. As is evident from their simulations, the confidence intervals have  improved coverage compared with standard methods. To our knowledge, the properties of their method has never been assessed for meta-analyses including fewer than five studies. In this manuscript, we propose a variation of the method by Henmi and Copas employing an improved estimator of the between-study heterogeneity, in particular when dealing with few studies only. In a simulation study, the proposed method is compared to several competitors. Overall, we found that our method outperforms the others in terms of coverage probabilities. In particular, an improvement compared with the proposal by Henmi and Copas is demonstrated. The work is motivated and illustrated by a systematic review and meta-analysis in paediatric immunosuppression following liver transplantations. \\

\end{abstract}

\section{INTRODUCTION} \label{sec:intro}

Systematic reviews aim to summarize all the available evidence relevant to a particular research question. If appropriate, 
the data from identified studies are quantitatively combined in a meta-analysis. If the true effect is the same in all studies to be combined in a meta-analysis, then the so-called common-effect or fixed-effect model is appropriate. In practical applications this assumption often appears to be too strict as some level of between-trial heterogeneity in the effects is suspected. Then the random-effects model is applied which treats the latent study-specific effects as random from an assumed distribution, often a normal distribution. Inference regarding the overall effect in the standard model relies on normality assumptions which apply if the number of trials and the trials themselves are sufficiently large. A number of estimators for the between-trial heterogeneity have been proposed \cite{VeronikiEtAl2015}; among them the standard approach by DerSimonian and Laird \cite{DerSimonianLaird86}. 

Recently the topic of meta-analysis of few studies, say 2 - 5, got more attention, since this case is very common in practice. With few studies, however, confidence intervals of the overall effect based on normal quantiles tend to be too short as they ignore the uncertainty in estimating the between-trial heterogeneity. As remedies, methods based on $t$-quantiles have been proposed \cite{FollmannProschan1999, HartungKnapp01a, HartungKnapp01b, SidikJonkman02}. With few studies only, however, they are often conservative and so long that they are uninformative \cite{RoeverEtAl2015}. Also various likelihood-based methods have recently been assessed in the specific situation of few studies and not found to be a solution to the problem \cite{SeideEtAl18}. Between-trial heterogeneity estimates mentioned above often result in zero \cite{FriedeEtAl2017a}, with the notable exception of the method proposed by Chung et al \cite{ChungEtAl2013}. Chung et al suggested the so-called Bayes modal (BM) estimator, which uses in a Bayesian framework a weakly informative prior for the between-trial heterogeneity to avoid zero estimates of the heterogeneity. Furthermore, a fully Bayesian approach to random-effects meta-analysis with weakly informative priors for the between-trial heterogeneity parameter has some advantages in this situation, since zero estimates are avoided as with the BM estimator and in addtition the uncertainty in estimating the heterogeneity is accounted for \cite{FriedeEtAl2017a, FriedeEtAl2017b}. Of course the Bayesian credible intervals would not necessarily have frequentist properties. Evaluating the operating characteristics in extensive simulation studies, it was found that the frequentist coverage probabilities are often above the nominal level with conservative choices of the prior for the between-trial heterogeneity \cite{FriedeEtAl2017a, FriedeEtAl2017b}. Bender et al\cite{BenderEtAl2018} recently provided an overview on the topic of meta-analyses with few studies.

In systematic reviews, relevant evidence is identified through systematic searches of literature databases. If all relevant studies would be published, this would be sufficient. However, this is not always the case. The problem was first described as the 'file drawer problem' \cite{Rosenthal79}.  Today various types of reporting biases are carefully distinguished including publication bias, time lag bias, citation bias and outcome reporting bias to name but a few. Studies might not be published at all for various reasons or only with a certain delay or in journals or languages that are more difficult to access (see e.g. Table 7.2.a in \cite{BoutronEtAl2019}). In the following we focus here on the aspect of publication bias. Prospective registration of clinical trials is one way to tackle this problem. It has become standard practice to search not only at least two electronic databases of the literature but also to search at least one registry for clinical studies such as clinicaltrials.gov. The idea would be to include unpublished studies in systematic reviews. However, access to the unpublished results is often challenging as it requires the cooperation of investigators, sponsors etc.  

A number of methods have been proposed over the years to deal with publication bias \cite{JinEtAl2015}. A popular way to interrogate data for publication bias is the visualization in form of a so-called funnel plot. In this scatter plot, each study contributes an estimate of an effect measure and its estimated standard error. The former is plotted on the x-axis, the latter on the y-axis. If no publication bias is present, we would expect the plot to be symmetric to the vertical line running through the average effect. Any absence of this symmetry might be interpreted as a signal that some form of reporting bias might be present. As this can be difficult to judge, formal hypothesis tests have been proposed (see e.g. \cite{EggerEtAl1997}). The problem with the visual inspection as well as with the formal tests is that they become more powerful with larger number of studies, but are less sensitive with few studies only. In the context of funnel plots, trim-and-fill methods have been proposed to correct the overall effect for potential publication bias \cite{DuvalTweedie2000}. Following an alternative approach, several sensitivity analysis methods have been suggested based on selection functions   describing the selective publication process \cite{CopasShih2000, CopasJackson2004, HenmiEtAl2007}. For instance, Copas and Jackson \cite{CopasJackson2004} investigated the maximum bias over all possible selection functions which satisfy the (fairly weak) condition that studies with smaller standard errors are at least as likely to be selected than studies with larger standard errors. Building on their work, Henmi et al \cite{HenmiEtAl2007} developed sensitivity analyses that, in contrast to the proposal by Copas and Jackson \cite{CopasJackson2004}, account for uncertainty in estimation. Again, the methods are not designed for the setting of few studies only.

Our work is motivated by a systematic review and meta-analysis of controlled clinical trials assessing efficacy and safety of Interleukin-2 receptor antagonists (IL-2RA) in children having undergone liver transplantations \cite{CrinsEtAl2014}, a rare surgical procedure in children. In total only six relevant studies were identified with little standardization with regard to the design of the studies implying some level of heterogeneity. Although the authors carefully checked for publication bias using standard techniques, it cannot be excluded that in particular some smaller studies were not published if they resulted in inconclusive treatment effects.

In contrast to the approaches to publication bias described above, Henmi and Copas \cite{HenmiCopas2010} proposed a method for random-effects meta-analysis that is robust to the selection of studies. They modified the DerSimonian-Laird (DL) confidence interval (\cite{DerSimonianLaird86}, (10) in \cite{HenmiCopas2010}) by replacing the random-effect estimator by the fixed-effect estimator of the overall effect and by replacing the normal quantiles by more accurate ones. The latter depends on the between-trial heterogeneity. The DL estimator is used in the computation of the quantiles. Therefore, with few studies this approach may not work well. In this paper, we propose a modification of the Henmi-Copas method by replacing the estimator of the between-study heterogeneity in the computation of the quantiles by the one developed by Chung et al \cite{ChungEtAl2013}. The properties of the new approach are assessed and compared to alternative methods including the Henmi-Copas approach and a proposal by Doi et al \cite{DoiEtAl2015} in Monte Carlo simulation studies considering in particular the case of few studies with and without publication bias. Our method is not conditional on having detected publication bias, e.g. in a funnel plot, since this would be very difficult with only few studies included in the meta-analysis. But it is robust to the selection of studies even with few studies as we will see below.

The manuscript is organized as follows. In the next section the new confidence interval for the overall effect is developed starting by introducing notation and reviewing 
the method of Henmi and Copas \cite{HenmiCopas2010}. The simulation study assessing the properties of the new confidence interval in comparison to existing 
methods is presented in Section \ref{sec:sims}. In Section \ref{sec:example} the proposed method is applied to the motivating example. We close with a brief discussion of 
our findings and their limitations.

\section{METHODS} \label{sec:meth}

Adopting the notation by Henmi and Copas \cite{HenmiCopas2010}, the true effect of an individual study $i$ out of $n$ independent studies is denoted by $\theta_i$. Estimates $y_i$ of the effects $\theta_i$ are observed with stand errors $\sigma_i$. Here we consider the normal-normal hierarchical model \textsl{(NNHM)}, which is the standard model for random-effects meta-analysis. In the \textsl{NNHM}, it is assumed that the $\theta_i$ are from a normal distribution with expectation $\theta$ and variance $\tau^2$, i.e. 
\begin{equation} \label{NNHM:par}
\theta_i \vert \theta, \tau \sim N(\theta,\tau^2),  \quad i=1,\dots, n .
\end{equation} 
Furthermore, the effect estimators $Y_i$ follow (at least approximately) a normal distribution with expectation $\theta_i$ and variance $\sigma_i^2$, i.e.
\begin{equation} \label{NNHM:est}
Y_i \vert \theta_i \sim N(\theta_i,\sigma_i^2), \quad i=1,\dots, n .
\end{equation}
From Equations (\ref{NNHM:par}) and (\ref{NNHM:est}) they follow the marginal model
\begin{equation} \label{MargModel}
Y_i \vert \theta,\tau \sim N(\theta,\sigma_i^2+\tau^2), \quad i=1,\dots, n .
\end{equation}
If the between-trial heterogeneity $\tau^2$ is 0, then the random-effects model reduces to the so-called fixed-effect or common-effect model. 

The focus of our study is inference regarding $\theta$, the overall effect. A standard method to construct an estimator and a $(1-\alpha)$ confidence interval for $\theta$ was proposed by DerSimonian and Laird \cite{DerSimonianLaird86} (DL). In short, the DL estimator of $\theta$ is given by
\begin{equation}\label{randomest}
   \hat{\theta}_{R} = \sum \hat{w_{i}}Y_{i}/ \sum \hat{w_{i}} \ ,
\end{equation}
where $\hat{w_{i}}=1/(\sigma_{i}^2+\hat \tau^2_{DL})$. Here, the DL estimator $\hat \tau^2_{DL}$ of the between-study heterogeneity $\tau^2$ is given by
\begin{equation}\label{tau2DL1}
   \hat \tau_{DL}^2 = \max\left\{0, \ \frac{Q - (n-1)} {\sum w_i - \sum w_i^2/\sum w_i}\right\} \ .
\end{equation}
The weights $w_i$ are the fixed effect weights (with $\tau^2=0$), which are $w_{i}=1/\sigma_{i}^{2}$. Furthermore, $Q$ is the so-called $Q$-statistic defined by
\begin{eqnarray}\label{Qstat}
   Q = \sum w_{i}(Y_{i} - \hat{\theta}_{F})^2 \ ,
\end{eqnarray}
where $\hat{\theta}_{F}$ is the fixed (or common) effect estimator of the overall effect with
\begin{equation}\label{fixedest}
    \hat{\theta}_{F}=\sum w_{i}Y_{i}/\sum w_{i} \ .
\end{equation}
If the estimator $\hat \tau_{DL}^2$ is assumed to be a fixed constant as the true value of $\tau^2$, then it holds that
\begin{eqnarray}\label{apivotDL}
   Z = \frac{\hat{\theta}_{R} - \theta}{1/\surd \sum \hat{w_{i}}} \sim N(0,1) \ .
\end{eqnarray}
This results in the DerSimonian-Laird $(1-\alpha) \%$ confidence interval (DL) for $\theta$, which is given by
\begin{equation}\label{confiDL}
   \left(\hat{\theta}_{R} - z_{(1-\alpha)/2}\frac{1}{\surd {\sum \hat{w_{i}}}} \ ,
         \hat{\theta}_{R} + z_{(1-\alpha)/2}\frac{1}{\surd {\sum \hat{w_{i}}}}\right) \ ,
\end{equation}
where $z_{\gamma}$ is the $\gamma$ quantile of the standard normal distribution. 
The assumption that the estimate $\hat \tau_{DL}^2$ is the true value of $\tau^2$ might be reasonable when the between-study 
heterogeneity can be estimated with high precision, i.e. when the number of studies included in the meta-analysis is large. In medical 
applications, however, this is frequently not the case. As noted by several authors, the application of the DL approach in meta-analyses 
with small to moderate numbers of studies results in coverage probabilities below the nominal level $1-\alpha$ \cite{FriedeEtAl2017a}.

Henmi and Copas \cite{HenmiCopas2010} tackled the two problems that (a) the distribution of the pivot statistic is quite 
different from the standard normal distribution when the number of studies $n$ is small, and (b) the estimators of $\theta$ are biased 
due to selective publication of smaller studies with less favourable results (publication bias). With respect to the latter they note that 
the common (or fixed) effect estimator $\hat \theta_F$ is more robust to publication bias than the random-effects estimator 
$\hat \theta_R$ simply because smaller studies, which are less likely to be published when their outcome is not favourable, have a 
smaller weight in the construction of $\hat \theta_F$ than in $\hat \theta_R$. To address the problem of the normal approximation 
they derive the distribution of the pivot statistic based on the fixed effect estimator under the random-effects model. 
More specifically, the variance of $\hat \theta_F$ is 
\begin{eqnarray}
   V (\tau^2)= \frac{\tau^2 \sum w_{i}^2 + \sum w_{i}}{(\sum w_{i})^2}.
\end{eqnarray}
The variance $V (\tau^2)$ can be estimated by plugging in $\hat \tau_{DL}^2$ for $\tau^2$. We denote this estimator of $V (\tau^2)$ by 
\begin{eqnarray}
   V (\hat \tau_{DL}^2)= \frac{\hat \tau_{DL}^2 \sum \hat w_{i}^2 + \sum \hat w_{i}}{(\sum \hat w_{i})^2}.
\end{eqnarray}
Recall that the weights $\hat{w_{i}}$ also depend on $\hat \tau^2_{DL}$. Hence, the pivot statistic $U$ is given by
\begin{eqnarray}\label{apivotNew}
   U = \frac{\hat{\theta}_{F} - \theta}{\sqrt{V (\hat \tau_{DL}^2)}}.
\end{eqnarray}
The point in the derivation of the distribution of $U$ by Henmi and Copas \cite{HenmiCopas2010} is to take into account the random variation of $\hat \tau^2_{DL}$ in addition to $\hat \theta_F$ as follows.

The distribution function of $U$ can be written as
\begin{eqnarray}\label{distU}
     P(U \le u) = \left\{\begin{array}{l}
                                     1 - \displaystyle\int_{u}^{\infty}P\left(\left.Q \le f^{-1}\left(\frac{r}{u}\right) \right| R = r\right)p_{R}(r)dr \ \ ({\rm if} \ u \ge 0) \\
                                     \displaystyle\int_{-\infty}^{u}P\left(\left.Q \le f^{-1}\left(\frac{r}{u}\right) \right| R = r\right)p_{R}(r)dr \ \ ({\rm if} \ u < 0)
                                \end{array}
                       \right.,
\end{eqnarray}
where the random variable $R$ and the function $f$ are defined by
\begin{eqnarray}
     R = \frac{\sum w_i(Y_i - \theta)}{\sqrt{\sum w_i}} \ {\rm and} \ f(Q) = \sqrt{\frac{\sum w_i^2\{Q - (n-1)\}}{(\sum w_i)^2 - \sum w_i^2} + 1},
\end{eqnarray}
respectively.
The function $p_R(r)$ is the probability density function of $R$, which is the normal density with mean zero and variance $1+\tau^2(\sum w_i^2/\sum w_i)$.
The conditional distribution of $Q$ given $R$, which is necessary to calculate the integral in (\ref{distU}), is a little complicated, but it is well approximated
by the gamma distribution whose mean and variance coincide with the exact conditional mean $M(R)$ and variance $V(R)$ of $Q$ given $R$, respectively
(see \cite{HenmiCopas2010} and its Appendix A for the explicit formulas of $M(R)$ and $V(R)$ and their derivation). 
Since both of the conditional mean $M(R)$ and variance $V(R)$ depend on the unknown true value of $\tau^2$ as does the variance of $R$, Henmi and 
Copas \cite{HenmiCopas2010} proposed the use of the DL estimator $\hat \tau_{DL}^2$ for $\tau^2$ again to approximate these quantities.
Under this setting, the (approximate) $\gamma$ quantile $u_{\gamma}$ of $U$ can be obtained by means of numerical integration and optimization 
(see Appendix B in \cite{HenmiCopas2010} for an implementation in R) and hence a $(1-\alpha)$ confidence interval for $\theta$ is given by 


\begin{eqnarray}\label{confiNew}
   \left(\hat{\theta}_{F} - u_{\alpha/2}\sqrt{V (\hat \tau_{DL}^2)}, \ \hat{\theta}_{F} + u_{\alpha/2}\sqrt{V (\hat \tau_{DL}^2)}\right).
\end{eqnarray}

In simulation studies, Henmi and Copas \cite{HenmiCopas2010} could show that their approach improves coverage probabilities as compared 
to standard procedures including the DL approach. With only few studies included in the meta-analysis, however, the performance is not satisfying. 
The poor performance of the method in this particular situation is caused (at least partly) by the use of the DL estimator $\hat \tau_{DL}^2$ 
{\it in the computation of the quantiles of the pivot statistic $U$} as above, since $\hat \tau_{DL}^2$ results frequently in zero estimates 
with few studies although the between-trial heterogeneity is positive, $\tau^2>0$. 

The use of weakly informative priors for the between-study heterogeneity to avoid zero estimates has been advocated for some time, whereas an 
uninformative, e.g. improper uniform, prior is used for the effect $\theta$ \cite{SpiegelhalterEtAl2004, Gelman2006}. A number of suggestions have 
been made on the choice of such weakly informative priors for $\tau$ including half-t \cite{Gelman2006} and half-normal distributions \cite{FriedeEtAl2017a}. 
Here we follow Chung et al \cite{ChungEtAl2013} who proposed to use a gamma distribution with shape $\eta$ and rate $\lambda$ as a prior for $\tau$, 
specifically $p(\tau) =  \lambda^{\eta}\tau^{\eta-1} e^{-\lambda \tau} / \Gamma(\eta)$ with gamma function $\Gamma(\eta)$. This choice means that 
the logarithm of the posterior of $\theta$ and $\tau$ is equal to the log likelihood plus a term depending only on $\tau$ but not $\theta$. Rather than 
using the mean or median of the posterior, Chung et al \cite{ChungEtAl2013} consider the mode, which can be computed by numerical optimization. 
This estimator of $\tau$ is referred to as the Bayes Modal (BM) estimator $\hat \tau_{BM}$. As default, Chung et al recommend to use $\alpha=2$ and 
$\lambda$ close to 0. The BM estimator $\hat \tau_{BM}$ can be interpreted as a penalized maximum likelihood (ML) estimator \cite{ChungEtAl2013}.

In this paper, we propose to replace the DL estimator $\hat \tau_{DL}^2$ in the computation of the quantiles of the pivot statistic $U$ by the BM estimator $\hat \tau^2_{BM}$. The choice of the BM estimator is motivated by its performance in comparison to other estimators in recent simulation studies (see e.g. Figures 2 and 3 in \cite{FriedeEtAl2017a}). The resulting $\gamma$ quantile is denoted by $u^{(BM)}_\gamma$. The $(1-\alpha)$ confidence interval for $\theta$ is then given by

\begin{eqnarray}\label{confiNewBM}
   \left(\hat{\theta}_{F} - u^{(BM)}_{\alpha/2}\sqrt{V (\hat \tau_{DL}^2)}, \ \hat{\theta}_{F} + u^{(BM)}_{\alpha/2}\sqrt{V (\hat \tau_{DL}^2)}\right).
\end{eqnarray}

In summary, our idea is that we still use the DL estimator $\hat \tau_{DL}^2$ in the construction of the pivot statistic $U$ given in (\ref{apivotNew}) 
in the same way as Henmi and Copas \cite{HenmiCopas2010}, but we use the BM estimator $\hat{\tau}^{2}_{BM}$ in the approximate calculation
for the distribution of $U$ instead of $\hat{\tau}^2_{DL}$. The reason for the use of the DL estimator $\hat{\tau}^2_{DL}$ in the construction of $U$ is that it is 
easier to calculate the distribution of the pivot statistic $U$, taking into account the effect of estimating $\tau^2$. However, the distribution of $U$ 
depends on the unknown true value of $\tau^2$ and it is necessary to use some estimate of $\tau^2$ to approximate the distribution of $U$. One possibility is 
to use the DL estimator $\hat{\tau}^2_{DL}$ again, which was done in \cite{HenmiCopas2010}, but it would be inaccurate unless the number of studies are 
sufficiently large. Hence, we propose the use of the BM estimator $\hat{\tau}^2_{BM}$ to improve the accuracy in estimating $\tau^2$ and in approximating the 
distribution of $U$, which we expect to lead the improvement of the coverage probabilities of the Henmi-Copas (HC) confidence interval (\ref{confiNew}). 
In the next section, by simulation studies, we show that the new confidence interval (\ref{confiNewBM}) actually improves the HC confidence interval 
(\ref{confiNew}) in coverage probability as well as the DL confidence interval (\ref{confiDL}) in both cases with and without publication bias, especially 
when the number of studies is small.

\section{SIMULATION STUDY} \label{sec:sims}

In order to compare the performance of the proposed approach with previously suggested procedures a Monte Carlo simulation study was conducted. As comparators the methods by Henmi and Copas \cite{HenmiCopas2010} (HC), Chung et al \cite{ChungEtAl2013} (BM), Doi et al \cite{DoiEtAl2015} (IVH) and DerSimonian and Laird \cite{DerSimonianLaird86} (DL) were included. The first one is known to be robust to publication bias to some extent, but its performance in meta-analyses with few studies only is unknown. The approach by Chung et al \cite{ChungEtAl2013} was developed for the scenario of few studies but might not be robust to publication bias. Doi et al \cite{DoiEtAl2015} proposed the inverse variance heterogeneity model. As with the HC approach, the interval is centred around an estimator assuming the common-effect model. Therefore, it might have attractive properties in settings with publication bias. In contrast to the HC approach, however, it is based on normal approximation. This approach was not included in recent method comparison studies \cite{VeronikiEtAl2019}. The DL approach was included here as it is often considered to be the standard approach to random-effects meta-analysis. The simulation model by Brockwell and Gordon \cite{BrockwellGordon2001} formed the basis for our simulation study. It was used in several recent simulation studies and therefore appeared to be a good choice. To account for publication bias, we used the same selection function (probability that a study with an outcome $y$ and associated standard error $\sigma$ is selected in the meta-analysis)
\begin{eqnarray}
     P(\mbox{selected} | y, \sigma) = \exp\left[-\beta\left\{\Phi\left(-\frac{y}{\sigma}\right)\right\}^{\gamma}\right]
\end{eqnarray}
as in \cite{HenmiCopas2010} with the same sets of the parameters $\beta$ and $\gamma$ for moderate and severe publication bias. Here, $\Phi$ is the cumulative distribution function of the standard normal distribution. Table \ref{tab_scenarios} summarizes the simulation scenarios considered. Per scenario $N=2,000$ simulation replications were run.

\begin{table} \begin{center}
\caption{Summary of the scenarios considered in the simulation study} \vspace{1mm} \label{tab_scenarios}
\begin{tabular}{ll}
\hline \hline
{\bf Parameter} & {\bf Values} \\
\hline
Treatment effect $\theta$ & $0.5$ \\
Between-trial heterogeneity $\tau^2$ & $0.05, 0.15, 0.25$ \\
Number of trials included in the meta-analysis $n$ & $3, 6, 9, 12, 15$ \\
Selection model & \\
\hspace{0.5cm} No publication bias & \\
\hspace{0.5cm} Moderate publication bias & $\beta=4$, $\gamma=3$ \\
\hspace{0.5cm} Severe publication bias & $\beta=4$, $\gamma= 1.5$ \\
\hline \hline
\end{tabular} \end{center}
\end{table}

\begin{figure} \begin{center}
{\footnotesize $\tau^2=0.05$ \hspace{3cm} $\tau^2=0.15$ \hspace{3cm} $\tau^2=0.25$} \\
\rotatebox{90}{\footnotesize \hspace{0.8cm} No publication bias}
\includegraphics[width=4.5cm]{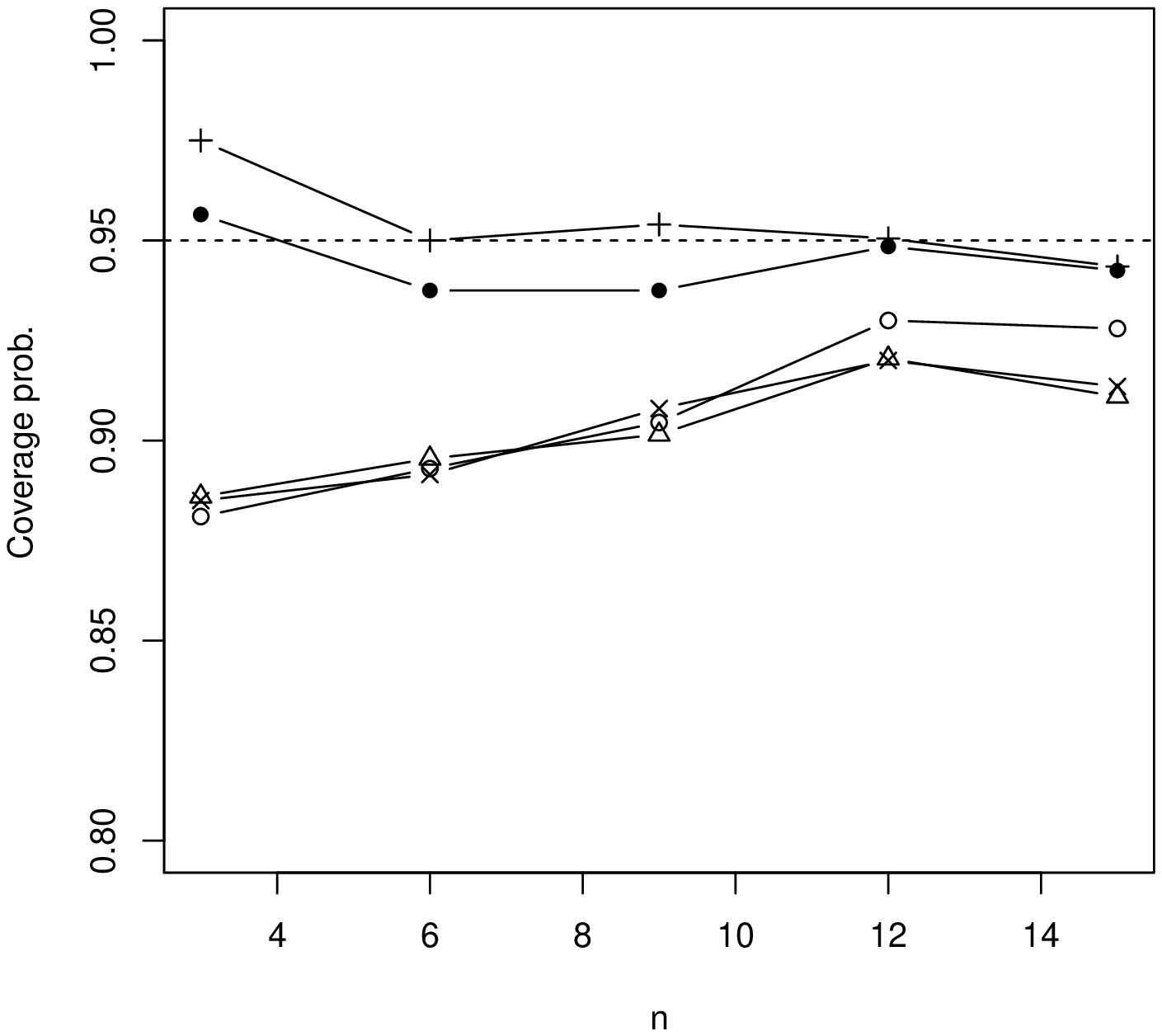}
\includegraphics[width=4.5cm]{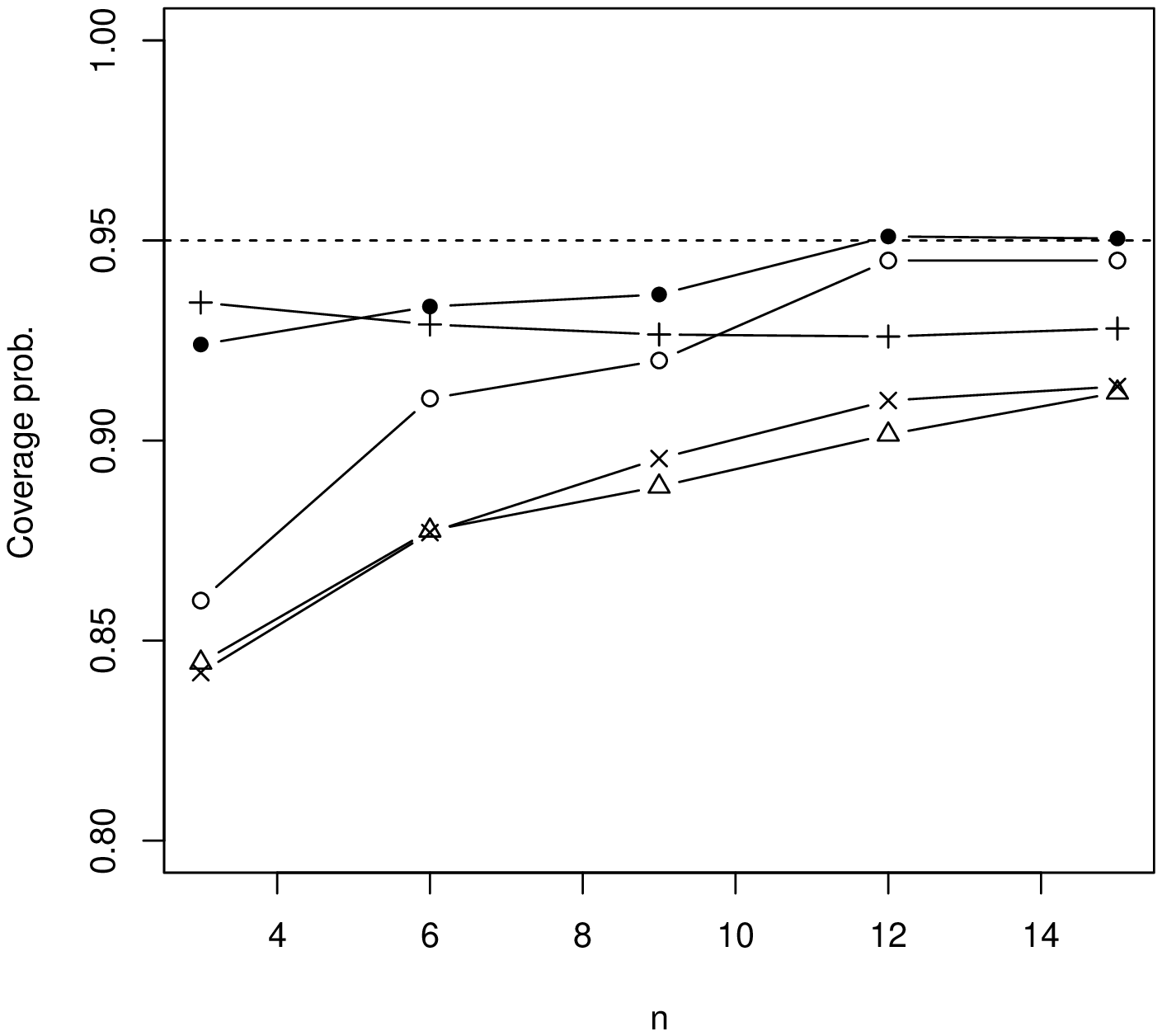}
\includegraphics[width=4.5cm]{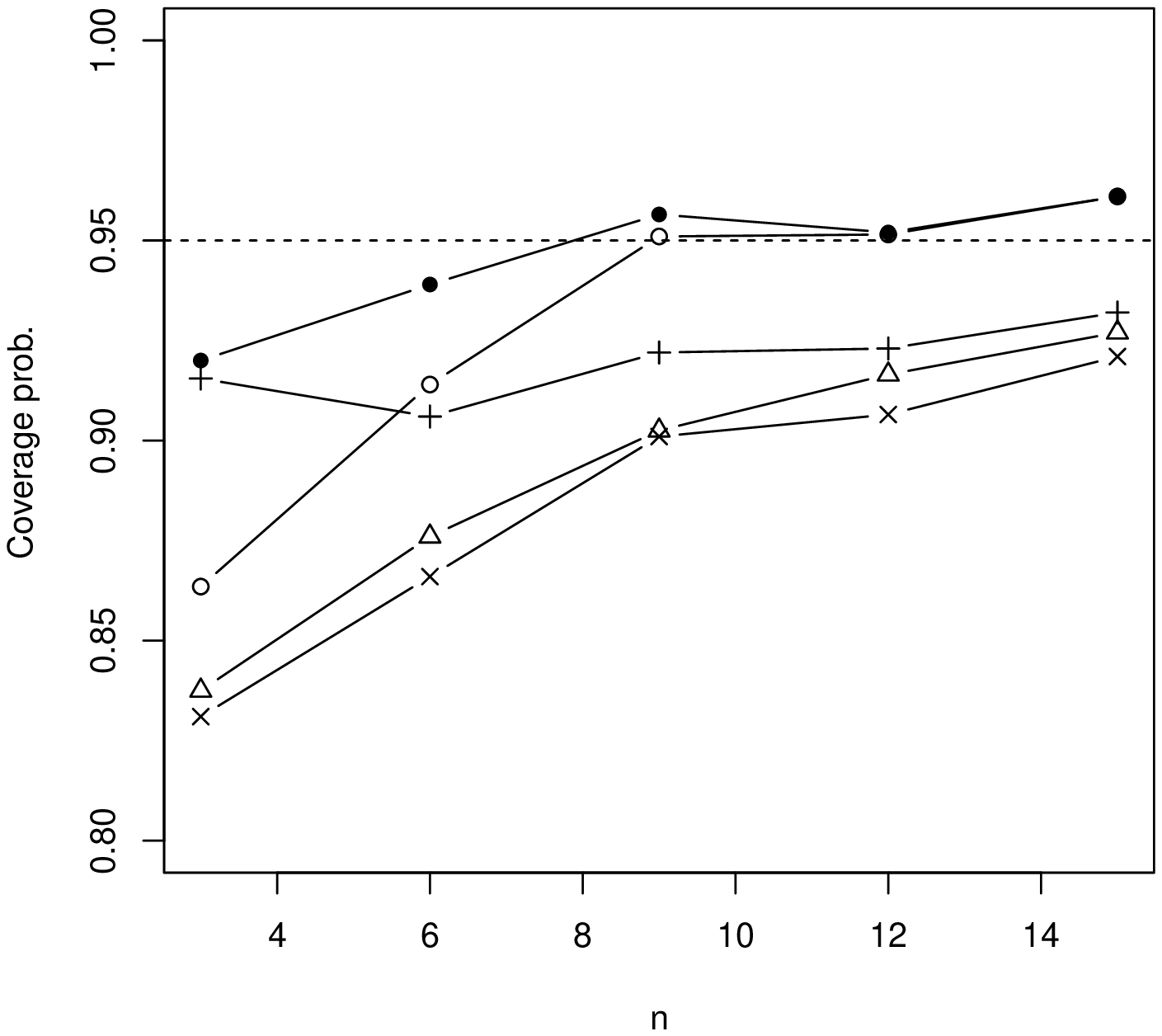} \\
\rotatebox{90}{\footnotesize \hspace{0.3cm} Moderate publication bias}
\includegraphics[width=4.5cm]{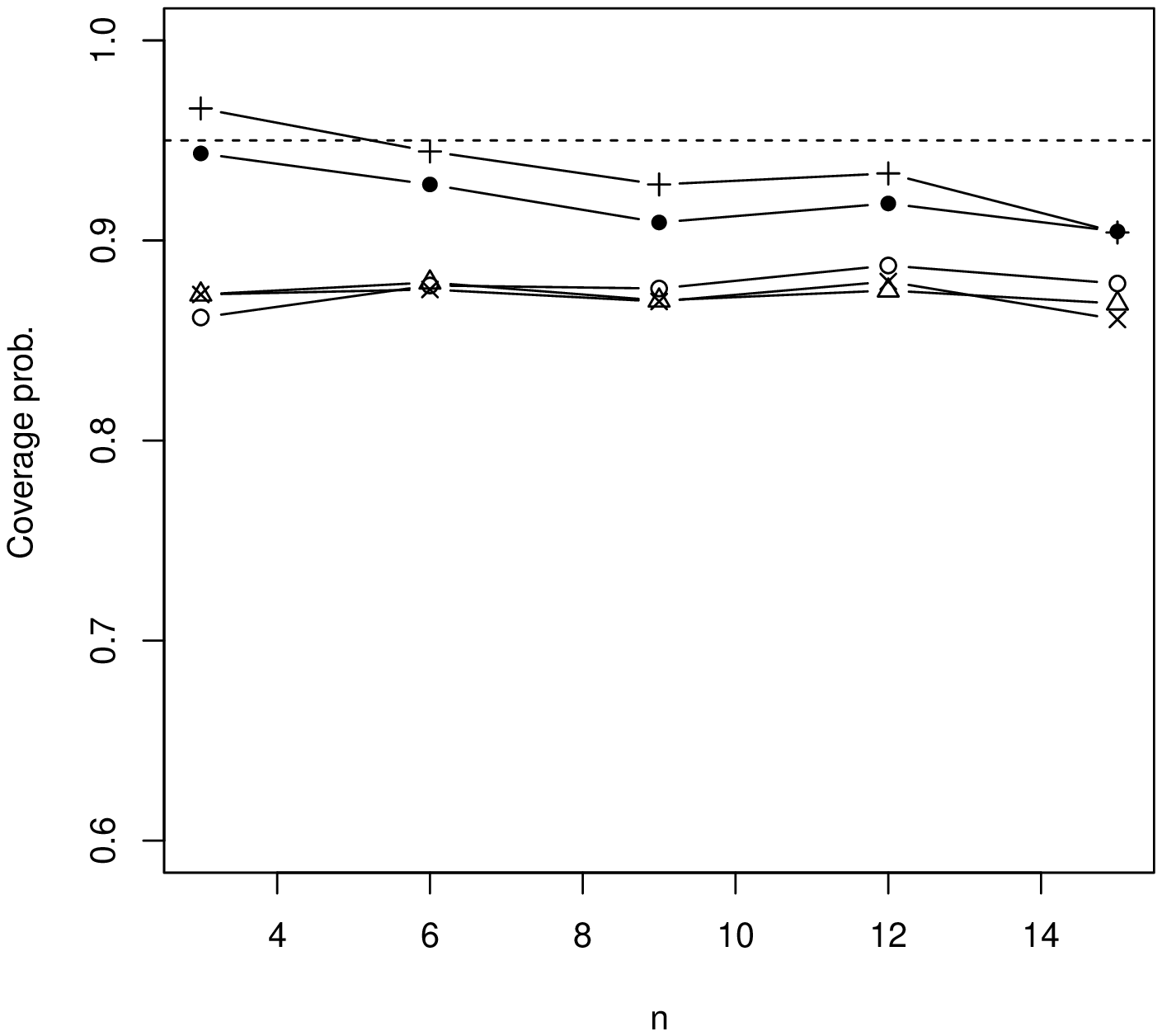}
\includegraphics[width=4.5cm]{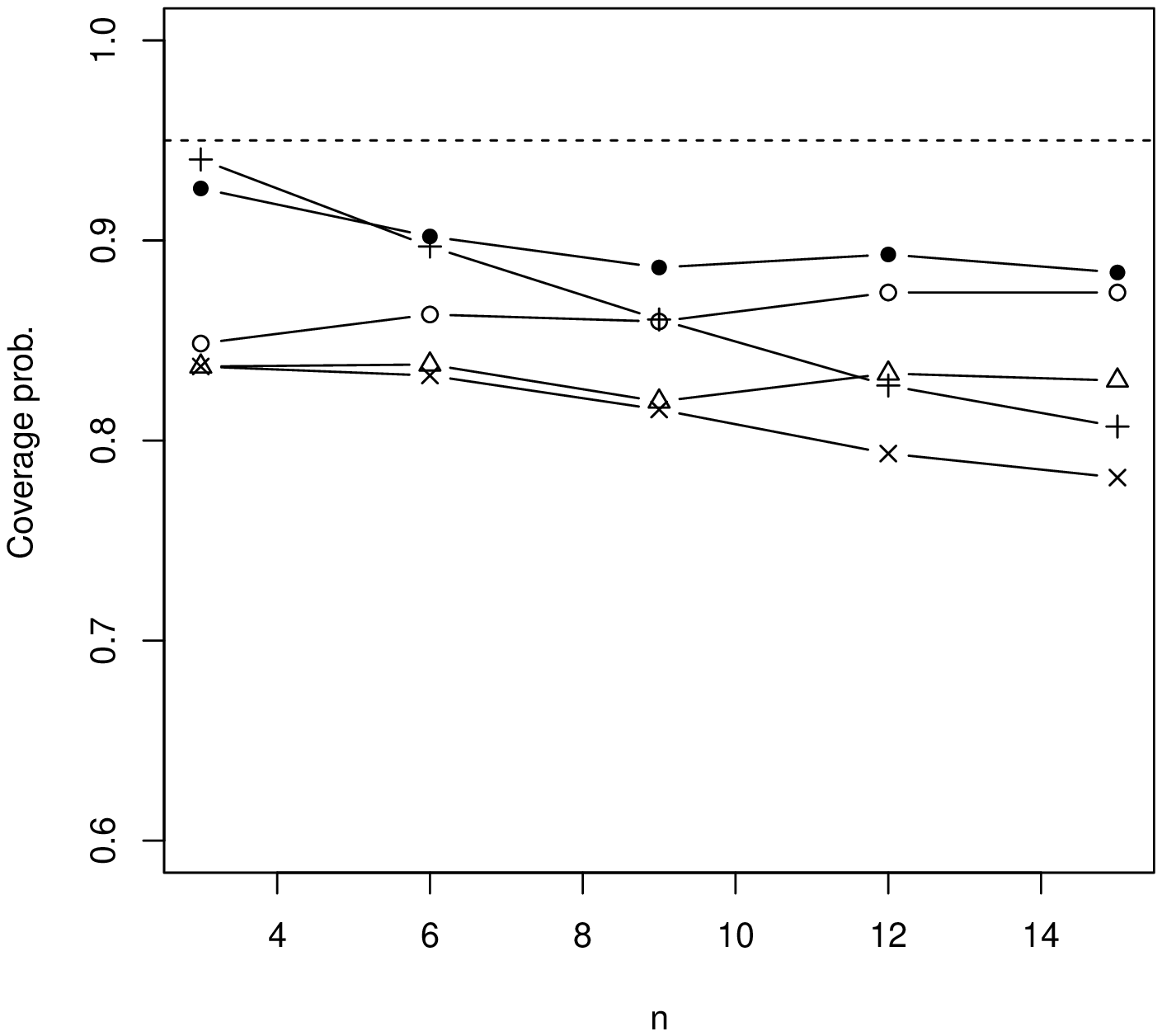}
\includegraphics[width=4.5cm]{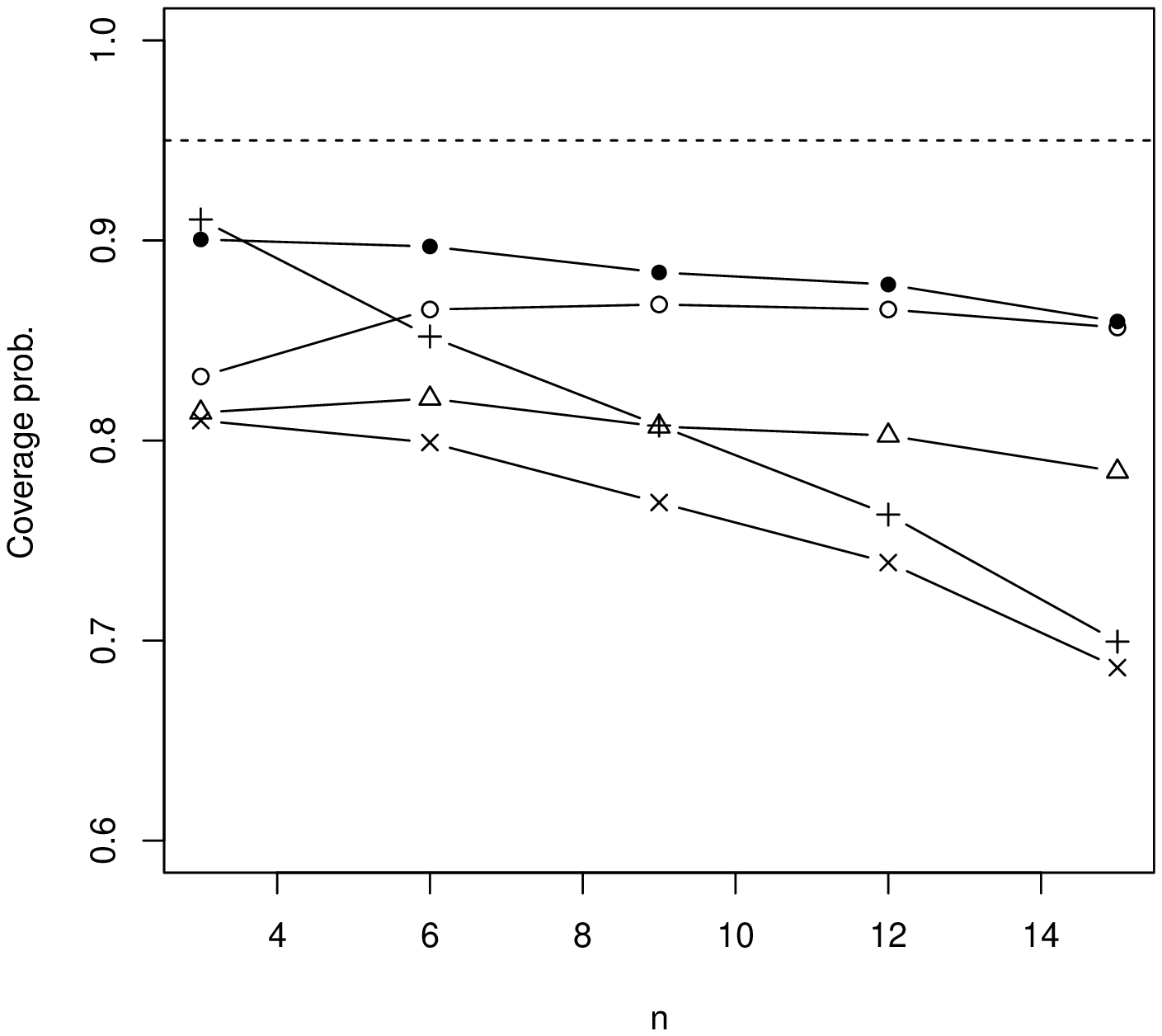} \\
\rotatebox{90}{\footnotesize \hspace{0.5cm} Severe publication bias}
\includegraphics[width=4.5cm]{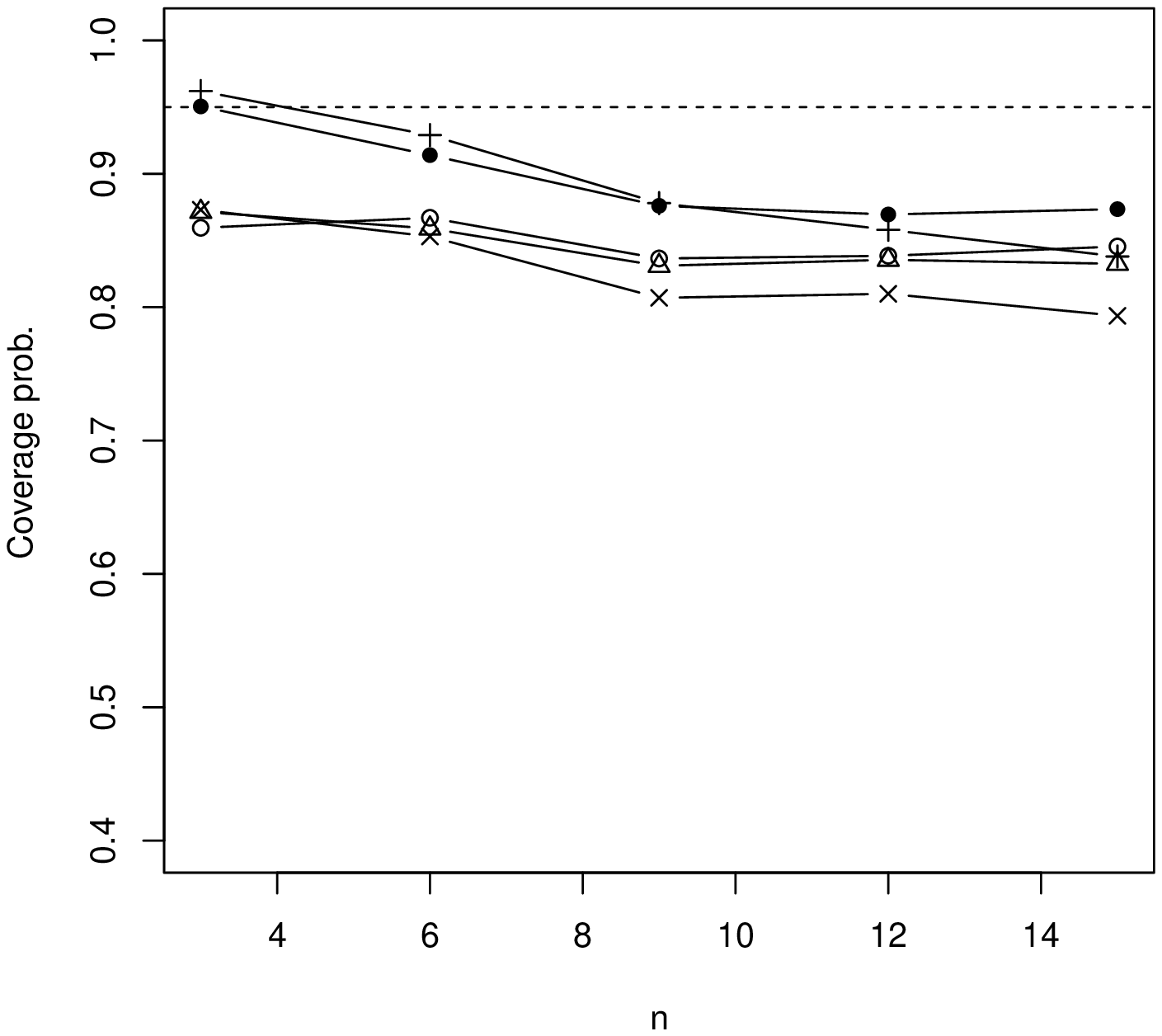}
\includegraphics[width=4.5cm]{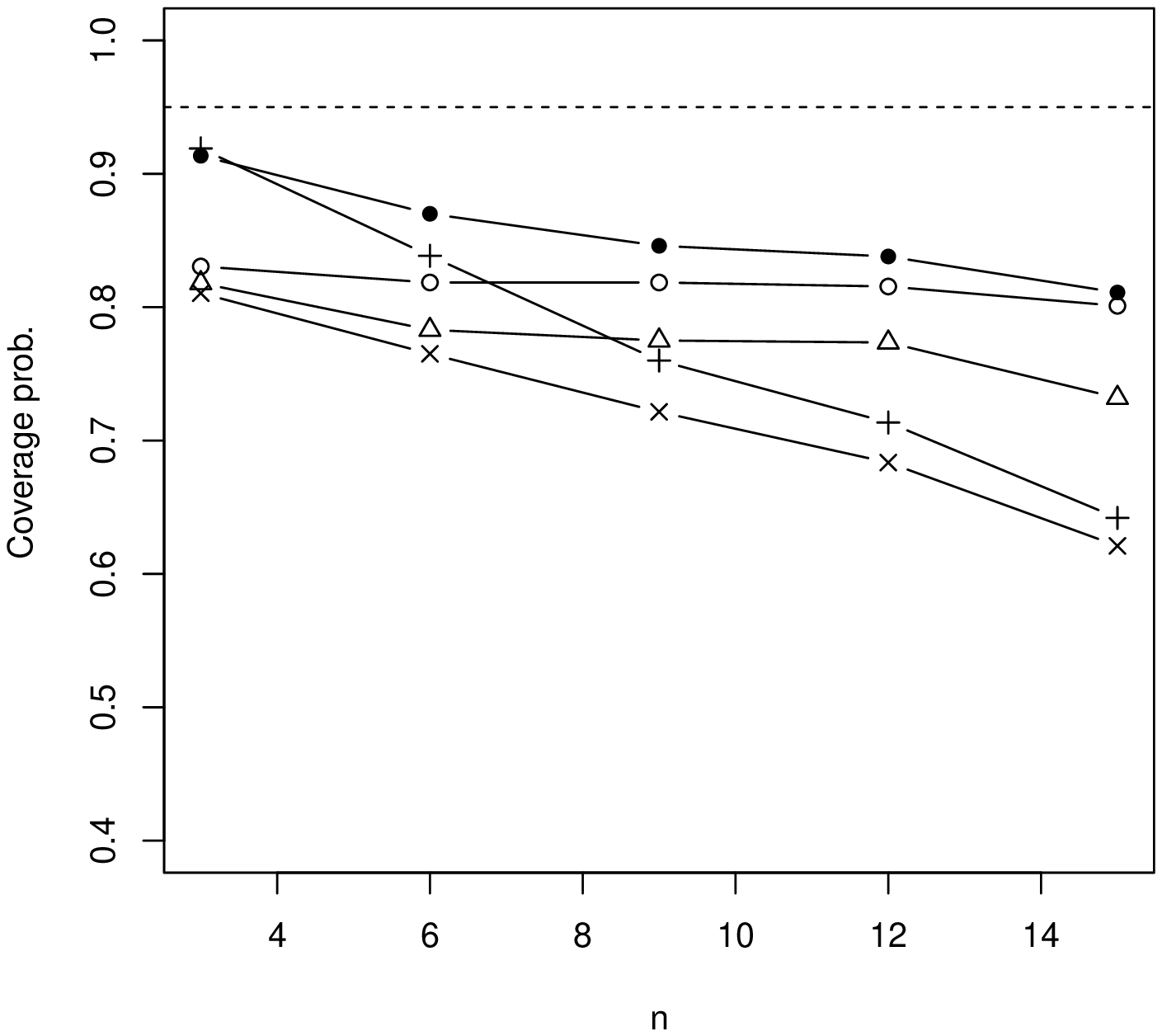}
\includegraphics[width=4.5cm]{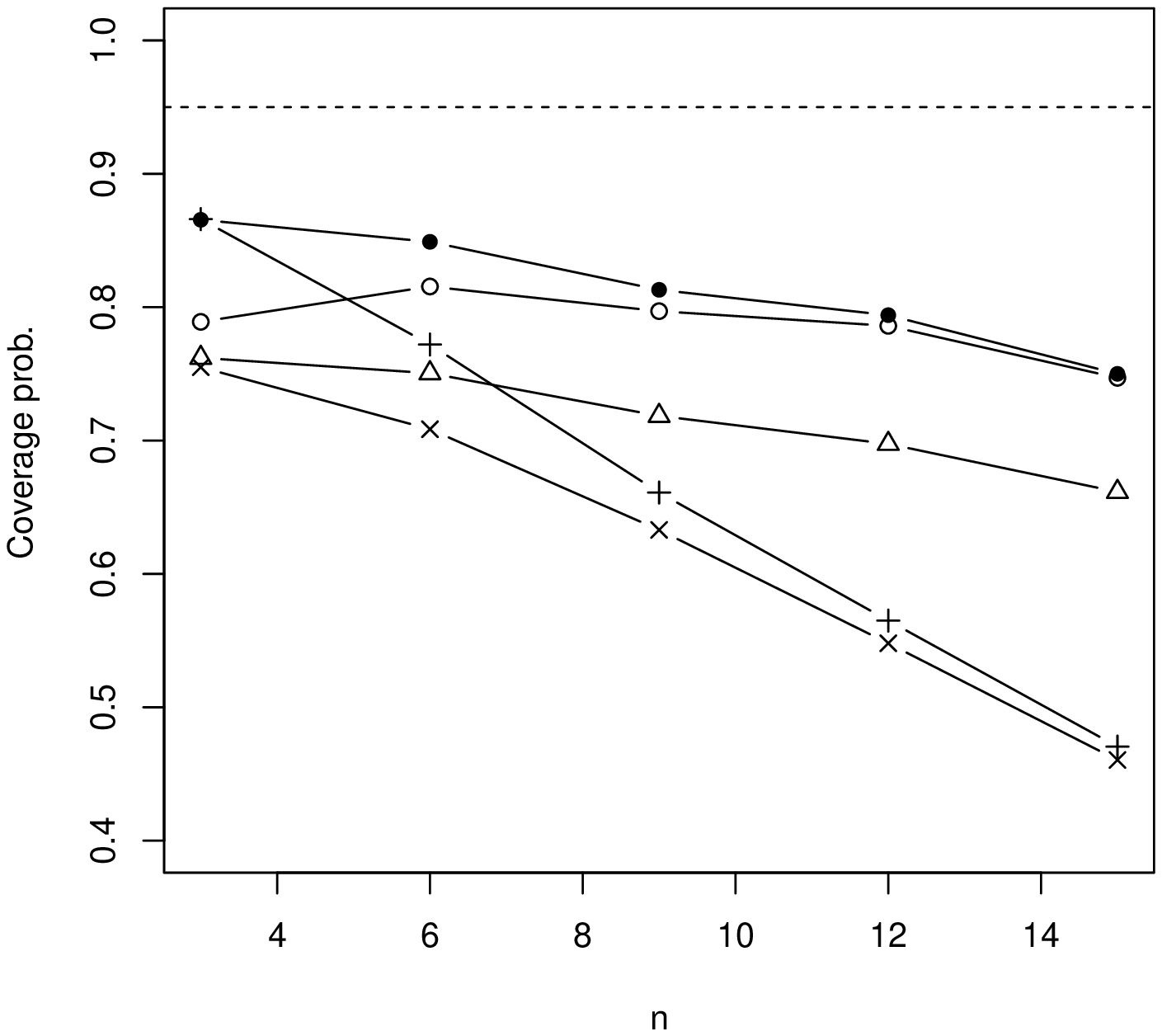} \\
\caption{Coverage probabilities of the various confidence intervals (circle: HC, cross: DL, dot: HC-BM, plus: BM, triangle: IVH) depending on the number of studies $n$ included in the meta-analysis for no, moderate and severe publication bias and for different degrees of between-trial heterogeneity $\tau^2 =  0.05, 0.15, 0.25$. } \label{fig_sims}
\end{center} \end{figure}

Figure \ref{fig_sims} presents the simulated coverage probabilities for the different confidence intervals in the various scenarios. In all scenarios considered, the proposed method performs at least as well as the HC method in terms of the coverage probability. With larger number of studies, say $n \ge 9$, and more pronounced between-trial heterogeneity, say $\tau^2 \ge 0.15$, the performance of both approaches is fairly similar. With smaller numbers of studies or only low levels of heterogeneity, however, there is a clear advantage for the new proposal as it improves the coverage probability considerably. In scenarios with few studies, $n=3$ or $n=6$, and only low levels of between-trial heterogeneity, $\tau^2=0.05$, the coverage probabilities of the BM approach are slightly higher than those of the proposed method. In the scenarios with publication bias, however, the coverage probabilities of the BM approach rapidly decrease well below the nominal level of 0.95 with increasing numbers of studies included in the meta-analysis and increasing levels of between-trial heterogeneity. Without publication bias, the coverage of the IVH interval is similar to the coverage of the DL interval, i.e. poor for small numbers of studies $n$ and closer to the nominal level for larger $n$. In the settings with publication bias the coverage probabilities of the IVH intervals are generally larger than those of the DL approach, in particular with more pronounced heterogeneity $\tau^2$ and larger numbers of studies $n$. However, the coverage probabilities are below those achieved by the HC and HC-BM approaches. Overall, the coverage probabilities of the proposed approach are closest to the nominal level whereas the coverages for the DL approach are well below the nominal level for several scenarios characterized by publication bias and small numbers of studies included in the meta-analysis.

In scenarios where different methods resulted in similar coverage probabilities close to the nominal level it is of interest to compare the length of the intervals obtained by these methods. Shorter intervals with the same coverage would of course be preferred. Table \ref{tab_cilength} gives the median interval lengths of the different confidence intervals for various levels of publication bias and heterogeneity $\tau^2$ as well as numbers of studies $n$ included in the meta-analysis. For instance, in the setting without publication bias and $n=3$ studies the median length of our proposed confidence interval (HC-BM interval) is 1.12, which is slightly larger than the median length of the BM intervals (1.15) although the coverage is 0.96 just below the coverage of the BM intervals (0.97). Similarly, with $n=6$ studies the median lengths of the HC-BM and BM intervals are 0.76 and 0.67, respectively. In scenarios where the coverages of the HC and HC-BM intervals are close, the median lengths of the intervals are similar again. In the scenarios without publication bias, where the coverages of the DL and IVH intervals are similar, the IVH intervals tend to be longer than the DL intervals when heterogeneity is present (i.e. $\tau>0$). This is expected since the common-effect estimator employed by the IVH interval is less efficient in this situation. As we have seen above, looking across all scenarios considered the HC-BM intervals achieve the best coverage. However, the price for this performance of coverage is paid by somewhat longer lengths when comparing to methods that achieve the desired coverage in particular situations but not in others.

\begin{table}[ht]
     \begin{center}
          \caption{Median lengths of the confidence intervals for various levels of publication bias and heterogeneity $\tau^2$ as well as numbers of studies $n$ included in the meta-analysis.} \vspace{1mm} \label{tab_cilength}
          \begin{tabular}{lccccccc} \hline \hline
          publication bias & $\tau^2$ & $n$ & DL & HC & HC-BM & BM & IVH \\ \hline 
          none &         0.05   & 3 & 0.815 & 0.878 & 1.134 & 1.109 & 0.830 \\ 
               &                & 6 & 0.577 & 0.678 & 0.756 & 0.683 & 0.605 \\ 
               &                & 9 & 0.466 & 0.577 & 0.606 & 0.524 & 0.496 \\ 
               &                & 12 & 0.412 & 0.517 & 0.532 & 0.450 & 0.443 \\ 
               &                & 15 & 0.362 & 0.458 & 0.465 & 0.391 & 0.395 \\ 
               &         0.15   & 3 & 1.005 & 1.291 & 1.536 & 1.269 & 1.051 \\ 
               &                & 6 & 0.748 & 1.145 & 1.194 & 0.832 & 0.837 \\ 
               &                & 9 & 0.634 & 0.957 & 0.966 & 0.672 & 0.735 \\ 
               &                & 12 & 0.557 & 0.817 & 0.816 & 0.584 & 0.663 \\ 
               &                & 15 & 0.505 & 0.726 & 0.724 & 0.522 & 0.608 \\ 
               &         0.25   & 3 & 1.174 & 1.778 & 2.007 & 1.421 & 1.246 \\ 
               &                & 6 & 0.892 & 1.488 & 1.515 & 0.971 & 1.039 \\ 
               &                & 9 & 0.749 & 1.208 & 1.207 & 0.796 & 0.915 \\ 
               &                & 12 & 0.658 & 1.004 & 0.998 & 0.686 & 0.815 \\ 
               &                & 15 & 0.591 & 0.885 & 0.880 & 0.615 & 0.753 \\ \hline
						
					moderate &   0.05   & 3 & 0.736 & 0.755 & 1.001 & 1.017 & 0.746 \\ 
                   &          & 6 & 0.503 & 0.533 & 0.620 & 0.614 & 0.515 \\ 
                   &          & 9 & 0.409 & 0.458 & 0.500 & 0.475 & 0.424 \\ 
                   &          & 12 & 0.358 & 0.407 & 0.436 & 0.407 & 0.373 \\ 
                   &          & 15 & 0.320 & 0.366 & 0.381 & 0.356 & 0.333 \\ 
                   &   0.15   & 3 & 0.860 & 0.963 & 1.207 & 1.132 & 0.884 \\ 
                   &          & 6 & 0.631 & 0.811 & 0.882 & 0.719 & 0.673 \\ 
                   &          & 9 & 0.518 & 0.707 & 0.721 & 0.566 & 0.572 \\ 
                   &          & 12 & 0.463 & 0.634 & 0.641 & 0.492 & 0.523 \\ 
                   &          & 15 & 0.421 & 0.567 & 0.568 & 0.438 & 0.481 \\ 
                   &   0.25   & 3 & 0.955 & 1.136 & 1.405 & 1.232 & 0.998 \\ 
                   &          & 6 & 0.721 & 1.053 & 1.099 & 0.798 & 0.791 \\ 
                   &          & 9 & 0.590 & 0.854 & 0.866 & 0.631 & 0.668 \\ 
                   &          & 12 & 0.537 & 0.782 & 0.784 & 0.562 & 0.636 \\ 
                   &          & 15 & 0.483 & 0.685 & 0.683 & 0.496 & 0.577 \\ \hline
									
          severe &    0.05    & 3 & 0.691 & 0.729 & 0.942 & 0.963 & 0.704 \\ 
                 &            & 6 & 0.477 & 0.510 & 0.598 & 0.587 & 0.489 \\ 
                 &            & 9 & 0.385 & 0.416 & 0.461 & 0.446 & 0.395 \\ 
                 &            & 12 & 0.334 & 0.371 & 0.394 & 0.379 & 0.345 \\ 
                 &            & 15 & 0.303 & 0.343 & 0.358 & 0.336 & 0.315 \\ 
                 &    0.15    & 3 & 0.827 & 0.916 & 1.162 & 1.083 & 0.849 \\ 
                 &            & 6 & 0.584 & 0.757 & 0.814 & 0.681 & 0.619 \\ 
                 &            & 9 & 0.495 & 0.674 & 0.691 & 0.540 & 0.541 \\ 
                 &            & 12 & 0.443 & 0.602 & 0.607 & 0.466 & 0.496 \\ 
                 &            & 15 & 0.403 & 0.539 & 0.542 & 0.421 & 0.456 \\ 
                 &    0.25    & 3 & 0.920 & 1.110 & 1.363 & 1.161 & 0.958 \\ 
                 &            & 6 & 0.683 & 0.988 & 1.031 & 0.759 & 0.744 \\ 
                 &            & 9 & 0.577 & 0.841 & 0.850 & 0.609 & 0.657 \\ 
                 &            & 12 & 0.509 & 0.726 & 0.728 & 0.528 & 0.595 \\ 
                 &            & 15 & 0.459 & 0.643 & 0.643 & 0.474 & 0.543 \\ \hline \hline
           \end{tabular}
     \end{center}
\end{table}

\section{APPLICATION TO A SYSTEMATIC REVIEW IN PAEDIATRIC LIVER TRANSPLANTATION} \label{sec:example}

Crins et al \cite{CrinsEtAl2014} report a systematic review and meta-analysis evaluating Interleukin-2 receptor antibodies
(IL-2RA) for immunosuppression in children who underwent liver transplantation. The authors identified a total of 
six controlled studies including two randomized trials. Given the heterogeneity in the designs of the studies, some 
between-study heterogeneity in the treatment effects can be expected. Although Crins et al \cite{CrinsEtAl2014} did not 
identify any publication bias by visual inspection of funnel plots and formal tests for asymmetry of these plots, this 
provides little reassurance that indeed no publication bias is present, since the number of studies is fairly small, 
which hinders the identification of publication bias in funnel plots or formal hypothesis tests. Therefore, there is a 
need for methods for random-effects meta-analyses robust to publication bias in this setting.

The endpoint acute rejections was reported in all six studies identified in the systematic review, whereas only three 
also reported the outcome steroid-resistant rejections. Table \ref{tab_example} summarizes the findings for both 
outcomes. These data were previously considered by Friede et al \cite{FriedeEtAl2017a} who applied several point 
estimators and confidence intervals of the overall effect including DL and BM to these. For acute rejections, DL 
and BM yielded log odds ratios (95\% confidence interval) of -1.59 (-2.21, -0.96) and -1.61 (-2.35, -0.87), respectively. 
The between-study heterogeneity was estimated as $\hat\tau^2_{DL}=0.16$ and $\hat\tau_{BM}^2=0.38$ with the DL 
and BM methods, respectively. The fixed-effect estimate of the overall effect is -1.56 smaller than the 
random-effects estimates. The HC interval given by (-2.24, -0.89) is centred around the fixed-effect estimate. 
The HC-BM interval proposed here is calculated as (-2.31, -0.82) which is considerably wider than the 
HC interval.

For steroid-resistant rejections, DL and BM resulted in a log odds ratio (95\% confidence interval) of -1.21 (-2.28, -0.15) 
and -1.32 (-2.78, 0.14), respectively. Whereas the DL method results in a statistically significant treatment difference, the effect is not statistically significant with the BM approach although the point estimate hints at a more 
pronounced treatment effect. This is explained by the larger between-study heterogeneity of $\hat\tau^2_{BM}=0.87$ 
with the BM method which compares to $\hat\tau_{DL}^2=0.14$ with the DL method. These compare to the fixed-effect 
estimate of -1.17 with 95\% confidence intervals of (-2.24, -0.09) and (-2.53,	0.20) for the HC and HC-BM
methods, respectively. Again, the fixed-effect estimate is smaller than the effects obtained from random-effects 
meta-analyses. Furthermore, the HC-BM confidence interval is wider than the HC interval. Here, this wider 
interval means that the effect is no longer statistically significant on the usual 5\% level.

\begin{table} \begin{center} 
\caption{Logarithm of the odds ratios (log OR) with standard errors (SE) for the endpoints acute rejection (AR) and 
steroid-resistant rejection (SRR) as reported in the systematic review by Crins et al \cite{CrinsEtAl2014}.} \vspace{1mm} \label{tab_example}
\begin{tabular}{lcccc}
\hline \hline
& \multicolumn{2}{c}{AR} & \multicolumn{2}{c}{SRR} \\
Study & log OR & (SE) & log OR & (SE) \\
\hline
Heffron (2003) & -2.31 & (0.60) & -2.00 & (0.91) \\
Gibelli (2004) & -0.46 & (0.56) & & \\
Schuller (2005) & -2.30 & (0.88) & & \\
Ganschow (2005) & -1.76 & (0.46) & -0.45 & (0.68) \\
Spada (2006) & -1.26 & (0.64) & & \\
Gras (2008) & -2.42 & (1.53) & -1.88 & (1.14) \\
\hline \hline
\end{tabular} 
\end{center} \end{table}

\section{DISCUSSION} \label{sec:discussion}

Meta-analyses of only a few studies are very common, but pose a number of challenges. These include the estimation of between-trial heterogeneity as well as the assessment of publication bias. Here we proposed a method that faces both challenges successfully. The confidence interval of the overall effect proposed by Henmi and Copas \cite{HenmiCopas2010} was improved by replacing the DerSimonian-Laird estimator by the Bayes Modal estimator of Chung et al \cite{ChungEtAl2013} in the computation of the quantiles to construct the confidence interval. The use of a weakly informative prior biases the Bayes Modal estimator away from zero. This resulted in larger quantiles, in particular in situations with few studies and only small to moderate levels of between-trial heterogeneity, which improved the coverage of the confidence intervals. 

There are a number of limitations. We focused on properties related to estimating the overall effect and did not consider other parameters such as the heterogeneity $\tau^2$ \cite{Jackson2006}. Furthermore, we refrained form investigating other selection functions, since Henmi and Copas state that their ``experience of working with other such models suggests that the extent of bias depends much more on the choice of selection parameters [\dots] than it does on the particular mathematical form of the selection function itself'' \cite{HenmiCopas2010}. Also, we did not include other comparators such as the Knapp-Hartung-Sidik-Jonkman approach \cite{HartungKnapp01a, HartungKnapp01b, SidikJonkman02}, since extensive comparisons were included in the paper by Henmi and Copas \cite{HenmiCopas2010} and also in more recent simulation studies \cite{FriedeEtAl2017a, FriedeEtAl2017b}.

The normal-normal hierarchical model considered here is a standard model for random-effects meta-analyses. This model is very general but not without limitations since effect estimates are modelled and not the data directly implying a two-step procedure. For instance, considering binary outcomes and treatment effects summarized by odds ratios Jackson et al \cite{JacksonEtAl2018} discuss six alternative generalised linear mixed models which are more efficient one-step procedures. Modelling the data directly can have particular benefits when dealing with rare events; see for example G\"unhan et al \cite{GunhanEtAl2019} or Gronsbell et al \cite{GronsbellEtAl2019}.

The approach taken here to improve the coverage of confidence intervals of the overall effect in pairwise meta-analysis might also be useful in more complex settings such as meta-regression or network meta-analysis. The exploration of such opportunities is out of the scope of this manuscript but subject of future research.

\section*{Highlights}
What is already known
\begin{itemize}
\item	Estimated overall effects from meta-analyses might be impacted by reporting bias
\item	A confidence interval for the overall effect has been proposed that is to some extent robust to the selection of studies
\end{itemize}

What is new
\begin{itemize}
\item	The performance of the robust confidence interval previously proposed is assessed in meta-analyses with few studies and found not to work well in this setting
\item	The approach is refined resulting in improved coverage probabilities of the confidence intervals in particular in meta-analyses with few studies
\end{itemize}

Potential impact for RSM readers outside the authors’ field
\begin{itemize}
\item	The refined approach is recommend for application in meta-analyses with few studies yielding more reliable results 
\end{itemize}

\section*{Data availability statement}
The data used in Section \ref{sec:example} are provided in Table \ref{tab_example}. Furthermore, they are given in the paper by Crins et al \cite{CrinsEtAl2014} and are also included in the R package {\tt bayesmeta} available from CRAN.

\section*{Acknowledgements}
The authors are grateful to Professor John Copas (Warwick) for discussions during his visit to Tokyo and Osaka in spring 2019.

\section*{ORCID}
Satoshi Hattori 0000-0001-5446-2305 \\
Tim Friede 0000-0001-5347-7441


\begin{thebibliography}{99}
\bibitem{VeronikiEtAl2015}Veroniki AA, Jackson D, Viechtbauer W, Bender R, Bowden J, Knapp G, {Ku\ss}  O, Higgins JPT, Langan D, Salanti G. Methods to estimate the between-study variance and its uncertainty in meta-analysis. \textit{Research Synthesis Methods} 2015; 7: 55--79.

\bibitem{DerSimonianLaird86}DerSimonian R, Laird N. Meta-analysis in clinical trials. \textit{Controlled Clinical Trials} 1986; 7: 177--188.

\bibitem{ChungEtAl2013}Chung Y, Rabe-Hesketh S, Choi I-H. Avoiding zero between-study variance estimates in random-effects meta-analysis. \textit{Statistics in Medicine} 2013; 32: 4071--4089.

\bibitem{FriedeEtAl2017a}Friede T, R\"over C, Wandel S, Neuenschwander B. Meta-analysis of few small studies in orphan diseases. \textit{Research Synthesis Methods} 2017; 8: 79--91.

\bibitem{FollmannProschan1999} Follmann DA, Proschan MA (1999). Valid inference in random effects meta-analysis. \textit{Biometrics} 1999; 55: 732--737.

\bibitem{HartungKnapp01a}Hartung J, Knapp G. On tests of the overall treatment effect in meta-analysis with normally distributed responses. \textit{Statistics in Medicine} 2001; 20: 1771--1782. 

\bibitem{HartungKnapp01b}Hartung J, Knapp G. A refined method for the meta-analysis of controlled clinical trials with binary outcome. \textit{Statistics in Medicine} 2001; 20: 3875--3889.

\bibitem{SidikJonkman02}Sidik K, Jonkman JN. A simple confidence interval for meta-analysis. \textit{Statistics in Medicine} 2002; 21: 3153--3159.

\bibitem{RoeverEtAl2015}R\"over C, Knapp G, Friede T. Hartung-Knapp-Sidik-Jonkman approach and its modification for random-effects meta-analysis with few studies. \textit{BMC Medical Research Methodology} 2015; 15: 99.

\bibitem{SeideEtAl18}Seide SE, R\"over C, Friede T. Likelihood-based meta-analysis with few studies: Empirical and simulation studies. \textit{BMC Medical Research Methodology} 2019; 19: 16.

\bibitem{FriedeEtAl2017b}Friede T, R\"over C, Wandel S, Neuenschwander B. Meta-analysis of two studies in the presence of heterogeneity with applications in rare diseases.  \textit{Biometrical Journal} 2017; 59: 658--671.

\bibitem{BenderEtAl2018}Bender R, Friede T, Koch A, {Ku\ss} O, Schlattmann P, Schwarzer G, Skipka G. Methods for evidence synthesis in the case of very few studies. \textit{Research Synthesis Methods} 2018; 9: 382--392.

\bibitem{Rosenthal79}Rosenthal R. The file drawer problem and tolerance for null results. \textit{Psychological Bulletin} 1979; 86: 638--641. 

\bibitem{BoutronEtAl2019}Boutron I, Page MJ, Higgins JPT, Altman DG, Lundh A, Hrobjartsson A. Chapter 7: Considering bias and conflicts of interest among the included studies. In: Higgins JPT, Thomas J, Chandler J, Cumpston M, Li T, Page MJ, Welch VA (editors). Cochrane Handbook for Systematic Reviews of Interventions version 6.0 (updated July 2019). Cochrane, 2019. Available from \url{www.training.cochrane.org/handbook}.

\bibitem{JinEtAl2015}Jin ZC, Zhou XH, He J. Statistical methods for dealing with publication bias in meta?analysis. \textit{Statistics in Medicine} 2015; 34: 343--360.

\bibitem{EggerEtAl1997}Egger M, Smith GD, Schneider M, Minder C. Bias in meta-analysis detected by a simple, graphical test. \textit{BMJ} 1997; 315: 629--634.

\bibitem{DuvalTweedie2000}Duval S, Tweedie R. A nonparametric "`trim and fill"' method of accounting for publication bias in meta-analysis. \textit{Journal of the American Statistical Association} 2000; 95: 89--98.

\bibitem{CopasShih2000}Copas J, Shih JQ. Meta-analysis, funnel plots and sensitivity analysis. \textit{Biostatistics} 2000; 1: 247--262.

\bibitem{CopasJackson2004}Copas J, Jackson D. A bound for publication bias based on the fraction of unpublished studies. \textit{Biometrics} 2004; 60: 146--153.

\bibitem{HenmiEtAl2007}Henmi M, Copas JB, Eguchi S. Confidence intervals and p-values for meta-analysis with publication bias. \textit{Biometrics} 2007; 63: 475--482.

\bibitem{CrinsEtAl2014}Crins ND, R\"over C, Goralczyk AD, Friede T. Interleukin-2 receptor antagonists for pediatric liver transplant recipients: a systematic review and meta-analysis of controlled studies. \textit{Pediatric Transplantation} 2014; 18: 839--850.

\bibitem{HenmiCopas2010}Henmi M, Copas JB. Confidence intervals for random effects meta-analysis and robustness to publication bias. \textit{Statistics in Medicine} 2010; 29: 2969--2983.

\bibitem{DoiEtAl2015}Doi SAR, Barendregt JJ, Khan S, Thalib L, Williams GM. Advances in the meta-analysis of heterogeneous clinical trials I: The inverse variance heterogeneity model. \textit{Contemporary Clinical Trials} 2015; 45: 130--138.

\bibitem{VeronikiEtAl2019}Veroniki AA, Jackson D, Bender R, Kuss O, Langan D, Higgins JPT, Knapp G, Salanti G. Methods to calculate uncertainty in the estimated overall effect size from a random-effects meta-analysis. \textit{Research Synthesis Methods} 2019; 10: 23--43.

\bibitem{SpiegelhalterEtAl2004}Spiegelhalter DJ, Abrams KR, Myles JP. \textit{Bayesian Approaches to Clinical Trials and Health-Care Evaluation}. Chichester: Wiley; 2004.

\bibitem{Gelman2006} Gelman A. Prior distributions for variance parameters in hierarchical models(Comment on Article by Browne and Draper). \textit{Bayesian Analysis} 2006; 1: 515--534.

\bibitem{BrockwellGordon2001}Brockwell SE, Gordon IR. A comparison of statistical methods for meta-analysis. \textit{Statistics in Medicine} 2001; 20: 825--840. 

\bibitem{JacksonEtAl2018}Jackson D, Law M, Stijnen T, Viechtbauer W, White IR. A comparison of 7 random-effects models for meta-analyses that estimate the summary odds ratio. \textit{Statistics in Medicine} 2018; 37: 1059--1085.

\bibitem{GunhanEtAl2019}G\"unhan BK, R\"over C, Friede T. Meta-analysis of few studies involving rare events. \textit{Research Synthesis Methods} 2019; (in press).

\bibitem{GronsbellEtAl2019}Gronsbell J, Hong C, Nie L, Lu Y, Tian L. Exact inference for the random-effect model for meta-analyses with rare events. \textit{Statistics in Medicine} 2019; (in press).

\bibitem{Jackson2006}Jackson D. The implications of publication bias for meta‐analysis' other parameter. \textit{Statistics in Medicine} 2006; 25: 2911--2921.

\end{thebibliography}
\end{document}